\documentclass[10pt,a4]{article} %
\usepackage[utf8]{inputenc}
\usepackage{amssymb}
\usepackage{amsmath}
\usepackage[top=3cm,bottom=2cm,left=2.0cm,right=2.0cm]{geometry}	
\usepackage[english]{babel}
\usepackage{pstricks}
\usepackage{verbatim}
\usepackage{graphicx}
\usepackage{multicol}
\usepackage{float}
\usepackage{hyperref}
\usepackage{natbib}
 \usepackage[bf,small]{caption}

\hypersetup
{
    pdfauthor={Júlio Caineta},
    pdfsubject={Dissertation for the degree of Master of Science of Mining and Geological Engineering},
    pdftitle={Applying Spatial Bootstrap and Bayesian Update in uncertainty assessment at oil reservoir appraisal stages},
    pdfkeywords={thesis, bootstrap espacial, update bayesiano, uncertainty, geostatistics}
    colorlinks=true
}
\begin{document}
\title{Applying Spatial Bootstrap and Bayesian Update in uncertainty assessment at oil reservoir appraisal stages}
\author{Júlio Alexandre Ramos Caineta\footnote{\textit{E-mail} julio.caineta@ist.utl.pt}\\
\textit{Under supervision of Amílcar de Oliveira Soares\footnote{\textit{E-mail} asoares@ist.utl.pt}}\\
\textit{Dept. Mining and Earth Resources, IST, Lisbon, Portugal}}
\date{December 2010}
\maketitle
\begin{abstract}
Geostatistical modeling of the reservoir intrinsic properties starts only with sparse data available. These estimates will depend largely on the number of wells and their location. The drilling costs are so high that they do not allow new wells to be placed for uncertainty assessment. Besides that difficulty, usual geostatistical models do not account for the uncertainty of conceptual models, which should be considered.

Spatial bootstrap is applied to assess the estimate reliability when resampling from original field is not an option. Considering different realities (conceptual models) and different scenarios (estimates), spatial bootstrapping applied with Bayesian update allows uncertainty assessment of the initial estimate and of the conceptual model.

In this work an approach is suggested to integrate both these techniques, resulting in a method to assess which models are more appropriate for a given scenario.
\end{abstract}
\bigskip
\begin{small}
\end{small}
\bigskip
\begin{multicols}{2}
\section{Introduction}

In the early stages of a petroleum reservoir development (exploration or appraisal stages) only sparse data is available, leading to inaccurate knowledge about that reservoir. Basically, these data comes from geologic maps, seismic surveys and a few wells. The latter has the highest costs. It is not so critical to get robust knowledge in terms of quantity, but of quality. The location of new wells (which accounts for a great economic risk) will deeply depend on that knowledge, hence being so important to measure, in some way its, accuracy, through what its commonly called uncertainty assessment.

In these stages, uncertainty is related to those few drilled wells and their location, to the stratigraphic interpretation obtained from core analysis and seismic data. Geostatistical modeling starts with only these sparse data. Possible geological scenarios are conceived gathering this information, but only part of it is retained to build conceptual models. These models are the starting point of the (re)definition of the drilling strategy (wells number and location), and therefore assessing uncertainty related to these models is fundamental for the field development. In this work a method is proposed to assess the uncertainty derived from these conceptual models.

The proposed approach to characterize conceptual models relies on two statistic techniques: spatial bootstrap, a resampling technique; and Bayesian update, an inference method based on Bayes' theorem. This approach is applied in a case study based on a Middle East reservoir, where its porosity is the analyzed variable.

We try to mimic the industry general methodology, starting with obtaining synthetic wells and building hypothetic scenarios (conceptual models) upon them. Once the first wells are drilled, the available data only allows a rough estimate of the given property (porosity, in this case). One way to assess the uncertainty of that estimate is to statistically analyze several realizations of the same estimator, i.e., new samples. In the petroleum context, new wells cannot be drilled from the real reservoir to assess uncertainty, since new samples would require drilling new wells, which accounts for a high economic risk (high uncertainty and high cost per drill). Bootstrap consists in a resampling technique which allows to measure the accuracy of an estimator discarding the need to acquire new realizations of that estimator. New samples are obtained by random sampling with replacement. This method can only be applied to independent variables, which is not convenient in a reservoir context where data are spatially correlated, as in earth sciences in general. When resampling wells, the correlation should be preserved as well as the redundancy between and along wells. Spatial bootstrap is then applied to overcome this limitation, which is accomplished by preserving the drilling strategy, i.e., the number and relative positions between wells are honored.

Similarly to the usual procedure followed in practice, an initial estimate is assumed as an hypothesis for each scenario. This value corresponds to a prior probability, which will then be transformed into a posterior probability through the inversion of the Bayes' theorem (Bayesian update), where spatial bootstrap is the key to calculate the likelihood. Different realities and different scenarios are integrated, resulting in an uncertainty model which considers the ignorance about what is assumed to be the reservoir structural model.

\section{Uncertainty in the early stages of a reservoir development}

The main goal of the exploration or appraisal stages is to find structures in the earth crust which are favorable for oil retention. Exploration consists in acquiring data (geological, geophysical or geochemical), in its analysis, and in the placement  of an initial exploration well (called \textit{wildcat}). This research can take place in some area not yet exploited or in the neighborhood of known fields, whereas in the second case they are called appraisal wells, serving the propose of finding structures associated to those fields. These initial wells are the only way one can validate the exploration conceptual model, providing the only practical evidence about a specific characteristic \citep{Gomes2007}.

Uncertainty is a measure of the ignorance we have about the physical phenomena which has occurred in a reservoir. Assessing reservoir properties and uncertainty through sparse appraisal data is a complex challenge. Particularly, the uncertainty assessment in a exploration or appraisal project should strive to answer the following two problems:
\begin{enumerate}
\item limited number of samples and limited knowledge about physical phenomena;
\item usually, the uncertainty related to the conceptual models is not taken into account.
\end{enumerate}
Various algorithms have been developed to calculate one unique best estimate of a global unknown value, however, assessing uncertainty requires the subjective definition of a randomization process. For instance, even when using stochastic simulation methods for quantifying the uncertainty degree of a geological model, through alternative and equiprobable representations, the uncertainty of the parameters, which have defined the geological model itself (e.g., variography), it is not quantified, because there was no place to consider other possibilities. A joint randomization of both the unknown variable and its estimate within a Bayesian framework, given a set of alternative but plausible geological scenarios, allows to account the uncertainty of the initial estimate and the geological model.

Bayesian update integrated with spatial bootstrap turns possible the uncertainty space creation for the conceptual models.

Although the developed methodology sees its major application in the context of the early stages of the petroleum reservoirs development, either by data scarcity or high associated costs (risk), its applicability might be extensible to other areas of knowledge.

\section{Method description}

The developed methodology is divided into two main groups, related to the concepts previously referred: spatial bootstrap and Bayesian update. Data will be produced in the first group and then processed in the second one.%

\subsection{Classes definition}

A specific estimator can be optimal to estimate lower values of a given property, but have a different behavior for middle or higher values. That possible trend can be verified through the definition of classes. These classes provide a range of values for the probabilities calculation, i.e., we are looking for the probability of the value of porosity which occurs between a specific range, instead of being equal to a specific value. That class can also relate to a range of interest, which in the petroleum context can correspond to ``be'' or ``not to be'' reservoir. From a general point of view, it is possible to know better the behavior of an estimator through the definition of $j$ classes $C_j$.

\subsection{Spatial bootstrap}
\label{sp}

Spatial bootstrap is a resampling technique, derived from the bootstrap method by \citet{Efron1977}. It is used to evaluate some specific statistic about a random variable $Z$. Spatial correlation between samples is preserved, while classical bootstrap does not account for spatial dependence.

In the spatial bootstrapping practice, as suggested by \citet{Journel1994}, stochastic simulations are run in order to obtain multiple realizations using the same set of parameters, which means that the structural model is not randomized. The same author proposed randomizing the simulation algorithm, retaining the \textit{real} value considered. In this approach, we use a spatial bootstrapping following, in some way, that suggestion, but instead of randomizing the simulation algorithm, we randomize the related variography (conceptual model). The introduced difference allows embedding different possible interpretations, given the sampling data.

Considering $k$ different variograms, we obtain different maps (each one with different spatial continuities), which will correspond to different realities $R_k$. In the context of this work, each resampling $A_m$ corresponds to the drilling of $m$ sets of $n$ synthetic wells on each one of the realities $R_k$.

Spatial bootstrapping procedure ends with this resampling. In the present work we propose a different approach. Stochastic simulations within spatial bootstrap are seen as realities, $R_k$, and the different scenarios, $S$, are defined by other simulations made upon the synthetic wells $A_m$. The number of simulations per each set $A_i$, with $i=1,\dots,m$, is the number of considered variograms, $k$.

With this procedure we can obtain the likelihood instead of using some random analytic distribution.

\subsection{Bayesian update}

\subsubsection{Prior probability}
\label{sss: prior}

Prior probability, $P(\phi\in C_j \mid S = s_k)$, corresponds to the probability of the occurrence of values of porosity within the class $j$, given the scenario $s_k$. Its definition is related to an initial hypothesis. That hypothesis can be derived from other informations previously acquired (e.g., seismic) and/or from expertise and available knowledge about some specific reservoir.

In this case, none of these are available, so we describe two ways to assign values to the prior probability. With the first one, the initial hypothesis comes from spatial bootstrap samplings. Frequentist probability is calculated for each one of the ``real'' maps. Going through each one of the $N$ sets of sampled wells, block by block, and considering the event of interest as the value of porosity being within the range defined for each class (Eq. \ref{eq: priori1}).
\begin{equation}
P(\phi\in C_j \mid S = s_k) = \frac{1}{N} \sum_{i=1}^{N} \phi_{A_i},
\label{eq: priori1}
\end{equation}
where $\phi_{A_i}$ is the frequentist probability related to the wells set $i$.

In the second way, successive values were assigned to the prior probability, depending on the real value at each ``real'' map, $\phi_{R_k}$, where $k$ corresponds to different conceptual models (Eq. \ref{eq: priori2}). Thereby, it is possible to verify prior probability behavior in a wider range of possible values.
\begin{equation}
P(\phi\in C_j \mid S = s_k) = \frac{P(\phi_{R_k} \in C_j \mid S=s_k)}{r},
\label{eq: priori2}
\end{equation}
with $r=0.1,0.2,\dots,3.0$. $r$ is a coefficient defined only to accomplish the propose of generating prior probability values in function of the ``real'' value.

\subsubsection{Evidence probability}

Bayesian definition of probability states that the probability of a specific event is a function of some evidence. That evidence corresponds to a sample or experimental information \citep{Ribeiro2001}. In this work, it is related to the $N$ simulations (scenarios) performed. The probability of the occurrence of porosity values with class $C_j$, in the simulated map $i$ of the scenario $s_k$, define the observed estimate $\phi^*_{S_{ik}}$. The probability of this experimental information is defined as the average value of $\phi^*_{S_{ik}}$, with $i=1,\dots,N$ (Eq. \ref{eq: sims}).
\begin{equation}
P(\phi^* \in C_j \mid S = s_k) = \frac{1}{N} \sum_{i=1}^{N} \phi^*_{S_{ik}}.
\label{eq: sims}
\end{equation}

\subsubsection{Likelihood}

Spatial bootstrapping procedure provides the likelihood of observing an estimate for a given proportion of porosity within the range $C_j$ and a geological scenario $s_k$ (Eq. \ref{eq: vero} shows its analytic definition). Likelihood is defined through the comparison between simulated images (scenarios) in the spatial bootstrap procedure, and the reality on which such procedure was carried on. The first step in this comparison consists in checking if the block $i$ of a specific ``real'' map is within the defined range for the class of interest. If so, we count the number of times that same block, in each simulated map, is also within the class. This procedure is repeated setting, in first place, one ``real'' map and, in second place, one scenario, which will be cycled over firstly.
\begin{equation}\label{eq: vero}
\begin{split}
P(\phi^* & \in C_j \mid \phi \in C_j, S = s_k) \\
& = P(\phi^* \in C_j, \phi \in C_j, S = s_k) \\
& \times \left[P(\phi \in C_j, S = s_k)\right]^{-1}.
\end{split}
\end{equation}

\subsubsection{Posterior probability}

Posterior probability (Eq. \ref{eq: update}) is the result obtained applying the Bayes' theorem. Its value refers to the probability of the occurrence of porosity values ($\phi$) within the class $C_j$, given an observed estimate $\phi^{*}$, in a specific scenario $s_{k}$.
\begin{equation}\label{eq: update}
\begin{split}
P(\phi & \in C_j \mid \phi^* \in C_j, S = s_k) \\
& = P(\phi^* \in C_j \mid \phi \in C_j, S = s_k) \\
& \times P(\phi \in C_j \mid S = s_k) \\
& \times \left[P(\phi^* \in C_j \mid S = s_k)\right]^{-1}.
\end{split}
\end{equation}
This probability corresponds to the initial hypothesis updated value (prior), and it is starting from which we will assess the model uncertainty.

\subsection{Application to the case study}

The data used comes from drilling and logging analysis performed in a Middle East reservoir. A synthetic reservoir was created (simulated) with that data, from which we start applying the developed methodology. In a real practical case, if there were well data available, this methodology would start with such data, thus it would not be necessary to use a simulated map.

The direct sequential simulation \citep{Soares2001} was used as the stochastic simulation algorithm. The parallelized version \citep{Nunes2010} was implemented in the workflow, due to its higher computational efficiency, which is one characteristic that should be taken into account when dealing with a large number of simulations.

\subsubsection{Considered variograms}

Three different variograms were set up, as shown in Table \ref{tab: amplitudes_r}. They were defined by changing the range, and they were used to generate three ``realities'' $R_{k}$ and three scenarios $S_{k}$, with $k=\text{G}, \text{M}, \text{P}$. These ranges were defined in function of a coefficient $c$, which corresponds to the approximate relation between the defined range in each direction and the corresponding size in the initial map. As an example, for the ``reality'' with the highest range (G), the relation with the initial map dimensions, in each direction, is approximately $1/2$, which is the maximum distance an experimental variogram should reach \citep{Soares2006}. The vertical direction was kept equal for all scenarios.

The spheric model was used in all cases. Different experimental variogram model choices can also be integrated into the $k$ definition.
\begin{table}[H]
\begin{small}
\begin{center}
\begin{tabular}{|c|c|c|c|c|}
\hline
\multicolumn{ 1}{|c|}{\textbf{$k$}} & \multicolumn{ 3}{|c|}{\textbf{Range in each direction}} & \multicolumn{ 1}{c|}{\textbf{$c$}} \\ \cline{ 2- 4}
\multicolumn{ 1}{|c|}{} & (90; 0) & (0; 0) & (0; 90) & \multicolumn{ 1}{c|}{} \\ \hline
G & 165 & 65 & 25 & $1/2$ \\ \hline
M & 110 & 45 & 25 & $1/3$ \\ \hline
P & 60 & 25 & 25 & $1/5$ \\ \hline
\end{tabular}
\end{center}
\caption{List of the considered variogram ranges.}
\label{tab: amplitudes_r}
\end{small}
\end{table}

\subsubsection{Classes definition}

Table \ref{tab: classes} shows the defined classes $C_{j}$, with $j=3Q, 1Q3Q, 1Q$. Ranges were defined using the quartiles from the initial map, $\phi_{I}$. In practice, these values should come from petrographic analysis.
\begin{table}[H]
\begin{small}
\begin{center}
\begin{tabular}{|l|c|}
\hline
\multicolumn{1}{|c|}{\textbf{Class}} & \textbf{Range} \\ \hline
$C_{3Q}$ & $[25.7612, \phi_{I_{max}}[$ \\
$C_{1Q3Q}$ & $[15.3437, 25.7612[$ \\
$C_{1Q}$ & $[\phi_{I_{min}},15.3437[$ \\ \hline
\end{tabular}
\end{center}
\caption{List of the defined porosity classes.}
\label{tab: classes}
\end{small}
\end{table}

\subsubsection{Spatial bootstrap}

\begin{figure*}[ht]
\begin{center}
\includegraphics[width=0.8\textwidth]{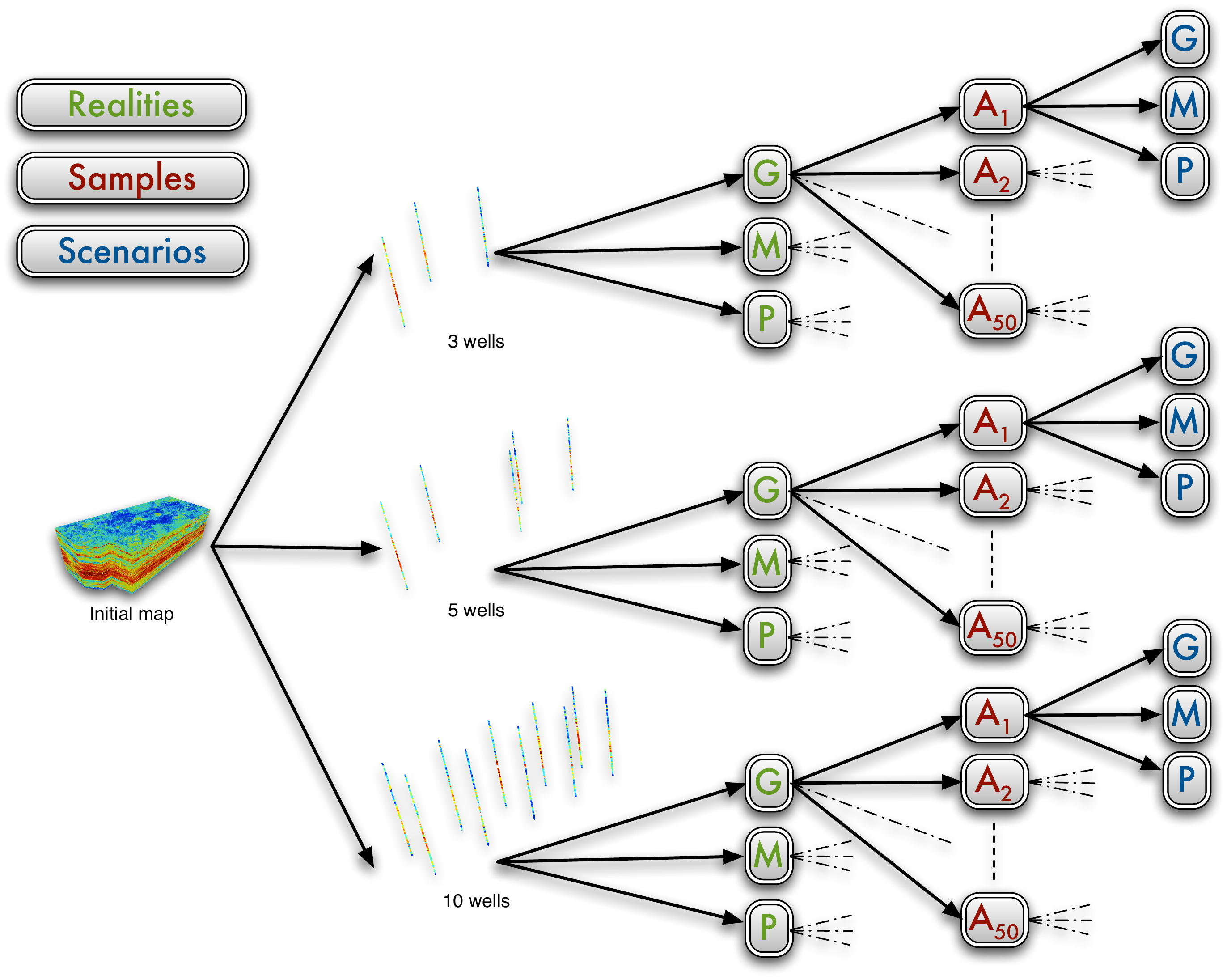}
\caption{Flowchart of spatial bootstrapping procedure applied to the case study.}
\label{fig: bootstrapflux_a}
\end{center}
\end{figure*}

Spatial bootstrap procedure was applied following the methodology described in Section \ref{sp}. Figure \ref{fig: bootstrapflux_a} illustrates the procedure followed from the ``objects'' point of view.
The initial data set consists in a porosity 3D map, originated from well data of a Middle East reservoir. From this map, three sets of synthetic wells were retrieved. These sets try to reproduce three different  drilling strategies. Table \ref{tab: coordenadas} shows all of their coordinates. The following procedure was followed for each one of those sets.
\begin{enumerate}
\item Three different variograms were defined (see Table \ref{tab: amplitudes_r}).
\item Three simulations were carried out, $R_{k}$, having each one associated to a different variogram $k$. Each one of these simulated maps corresponds to a different reality, resulting in three realities, distinguished by having different spatial continuities.
\item Fifty sets of wells were randomly sampled, $A_i$, with $i=1,\dots,50$. Each one of these sets honors the drilling strategy initially defined, $W$, i.e., the number of wells and their relative position are preserved.
\item Three new simulations were carried out for each one of the sampled sets, also using different variograms. Each one of these simulations corresponds to a possible scenario $S_{k}$, which matches with what is known about some specific reality $R_{k}$, i.e, the wells $A_{i}$.
\end{enumerate}
\vspace{-5mm}

\begin{table}[H]
\begin{small}
\begin{center}
\begin{tabular}{llllrr}
\multicolumn{ 6}{c}{\textbf{Number of wells}} \\ \cline{2-5}
\multicolumn{ 2}{c|}{\textbf{3}} & \multicolumn{ 2}{c|}{\textbf{5}} & \multicolumn{ 2}{c}{\textbf{10}} \\  \hline
\multicolumn{1}{c}{$x$} & \multicolumn{1}{c|}{$y$} & \multicolumn{1}{c}{$x$} & \multicolumn{1}{c|}{$y$} & \multicolumn{1}{c}{$x$} & \multicolumn{1}{c}{$y$} \\ 
\multicolumn{1}{r}{100} & \multicolumn{1}{r|}{40} & \multicolumn{1}{r}{60} & \multicolumn{1}{r|}{50} & 65 & 55 \\ 
\multicolumn{1}{r}{170} & \multicolumn{1}{r|}{90} & \multicolumn{1}{r}{130} & \multicolumn{1}{r|}{100} & 65 & 100 \\ 
\multicolumn{1}{r}{230} & \multicolumn{1}{r|}{60} & \multicolumn{1}{r}{200} & \multicolumn{1}{r|}{40} & 120 & 55 \\ 
 &  & \multicolumn{1}{r}{250} & \multicolumn{1}{r|}{90} & 120 & 100 \\ 
 &  & \multicolumn{1}{r}{310} & \multicolumn{1}{r|}{60} & 175 & 55 \\ 
 &  &  &  & 175 & 100 \\ 
 &  &  &  & 230 & 55 \\ 
 &  &  &  & 230 & 100 \\ 
 &  &  &  & 285 & 55 \\ 
 &  &  &  & 285 & 100 \\ 
\end{tabular}
\end{center}
\caption{Coordinates of the three sets of synthetic wells (in number of blocks).}
\label{tab: coordenadas}
\end{small}
\end{table}
\vspace{-4mm}
The number of wells in each synthetic set was chosen with the goal of approximating to real situations (reduced number of wells). It also serves the purpose of testing the methodology in different situations.
Overall, 1350 simulations were executed (3 drilling strategies $\times$ 3 realities $\times$ 50 samples $\times$ 3 scenarios).

\subsubsection{Bayesian update}

All probabilities necessary for the Bayesian update were calculated using algorithms developed during this work. Grids are traversed block by block and probabilities are calculated as described before, for each class $C_{j}$.

Posterior probabilities correspond to a new estimate of the proportion of porosity values within each class. These values are not conclusive by themselves, and the results give three possible values for each reality. It is necessary to assess which one of those estimates is the best one, i.e., which one is related to a lower uncertainty. An approach to reach out that goal is presented in the next section.

\subsection{Sensitivity analysis}
\label{ss: sensibilidade}

The models reliability assessment was achieved through a comparison between probabilities of the  ``real'' maps,  $P(\phi_{R_k} \in C_j \mid S = s_k)$, and the posterior probabilities, $P(\phi \in C_j \mid \phi^* \in C_j, S = s_k)$. Thus, it becomes possible to sort models by ranking. This ranking corresponds to a deviation, normalized to a percentage (Eq. \ref{eq: desvio}).
\end{multicols}
\par\noindent\rule{\dimexpr(0.5\textwidth-0.5\columnsep-0.4pt)}{0.4pt}%
\rule{0.4pt}{6pt}
\begin{equation}
D_{R1} = \frac{| P(\phi \in C_j \mid \phi^* \in C_j, S = s_k) - P(\phi_{R_k} \in C_j \mid S = s_k) |}{P(\phi_{R_k} \in C_j \mid S = s_k)}.
\label{eq: desvio}
\end{equation}
\begin{multicols}{2}
This deviation is no more than the absolute value of the difference between posterior probability and the ``real'' value, expressed as a percentage. High values of $D_{R1}$ indicate a higher difference between the final estimate (posterior probability) and the ``real'' value, also relating to a higher uncertainty. Lower values indicate the opposite, which is a better result. These values are more easily comparable between the same real map $R_{k}$, since the normalization is made with different values for different realities. There are three final values for each real map, which correspond to the three simulated scenarios.

Considering the assignment of prior values by the second hypothesis described in Section \ref{sss: prior}, Eq. \ref{eq: desvio} can be written replacing posterior probability by equation \ref{eq: update} and prior probability by Eq. \ref{eq: priori2}, resulting in the deviation $D_{R2}$ (Eq. \ref{eq: desvio2}),
\end{multicols}
\par\noindent\rule{\dimexpr(0.5\textwidth-0.5\columnsep-0.4pt)}{0.4pt}%
\rule{0.4pt}{6pt}
\begin{equation}
D_{R2} = \frac{ \left | \frac{P(\phi^* \in C_j \mid \phi \in C_j, S = s_k) \times P(\phi_{R_k} \in C_j \mid S = s_k)}{P(\phi^* \in C_j \mid S = s_k) \times r} - P(\phi_{R_k} \in C_j \mid S = s_k) \right |}{P(\phi_{R_k} \in C_j \mid S = s_k)} \times r,
\label{eq: desvio2}
\end{equation}
\vspace{\belowdisplayskip}\hfill\rule[-6pt]{0.4pt}{6.4pt}%
\rule{\dimexpr(0.5\textwidth-0.5\columnsep-1pt)}{0.4pt}
\begin{multicols}{2}
\noindent with $r=0.1,0.2,\dots,3.0$. Interpreting $D_{R2}$ values is the same as for $D_{R1}$, where the only difference is the range of prior values considered. One should consider, in such interpretation, the effect of similitude between images, i.e., likelihood. Very similar images will have high values of likelihood, which, in turn, tend to higher posterior probability values. Those updated values will be higher, thus increasing deviation. It results in high deviations in images which are similar by default. This leads to a difficult comparison between deviations related to different ``real'' maps, namely, between images generated with different number of wells, since the similitude between the real image and scenarios tends to be higher when the number of wells put into the simulation is also higher.

\begin{figure}[H]
\begin{center}
\includegraphics[width=0.5\textwidth]{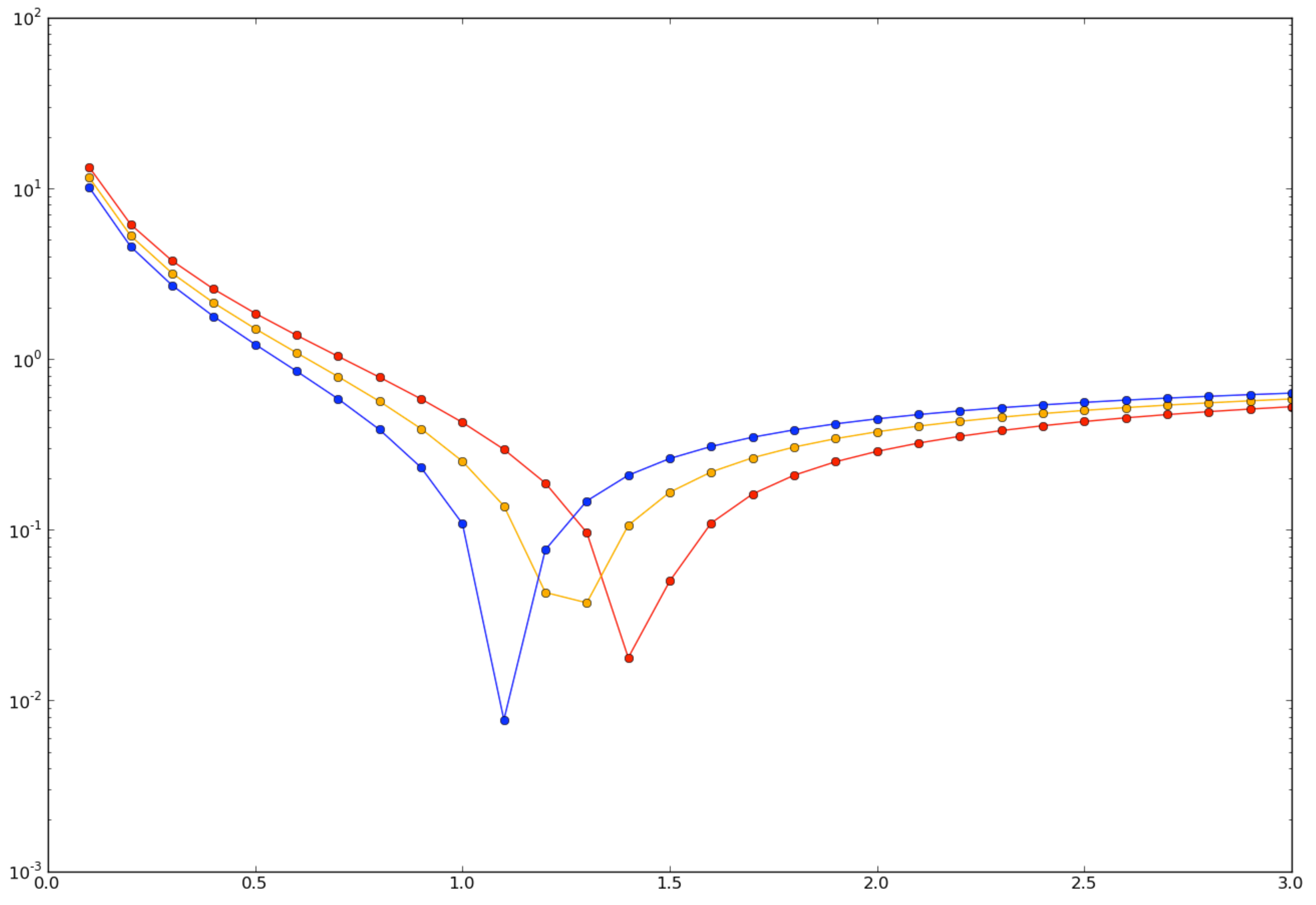}
\caption{Curves for model selection (red: $S_G$, yellow: $S_M$, blue: $S_P$) calculated from deviation $D_{R2}$, y axis (in logarithmic scale), and from coefficient $r$, x axis. Example shown for class $C_{3Q}$.}
\label{fig: desvio}
\end{center}
\end{figure}

Equation \ref{eq: desvio2} was applied to all realities and scenarios, resulting in 27 graphics, with 3 curves in each one. Figure \ref{fig: desvio} shows one of those graphics as an example. In this example, it is noticeable that there is an interval where the scenario $G$ has a lower deviation, other interval where is the scenario $M$ having a lower deviation, and yet another interval where scenario $P$ is the one with a lower deviation. This means that a different optimal model is obtained depending on the initial hypothesis (prior probability).

It was verified that all the curves have a similar behavior even between different real maps. The minimum value for $D_{R2}$ and $r$ values that define the intervals where each model is optimal changes, but the aspect and ordering relation are retained for most cases. Finding a mathematical relation between all these curves and building a function that defines them, which is far beyond the objectives of this work, would be helpful to find the optimal model expeditiously.
\vspace{-9mm}

\section{Conclusions}

\subsection{Summary}

We presented a methodology for uncertainty assessment in the early stages of a petroleum reservoir development. It can be said that in these stages there are two common difficulties: (a) available data are sparse and making decisions under such conditions has a high associated risk; (b) conceptual models are usually assumed as exempt of uncertainty, although it should be considered. This methodology incorporates two statistic techniques: spatial bootstrap and Bayesian update. Generally, it can be stated that spatial bootstrap helps minimizing difficulty (a) and Bayesian update helps minimizing difficulty (b).

The employed techniques were implemented with a few different approaches, seeking to improve its responses:
\begin{description}
\item[Spatial bootstrap] In relation to the methodology suggested by \citet{Journel1994}, two modifications were added regarding the way the conceptual models are randomized, or, in other words, their uncertainty assessment. We consider realities with different structural models (variograms) which are compared, each one of them, with a different set of scenarios (see Section \ref{sp}).

\item[Bayesian update] The presented method to calculate likelihood does not account for global proportions, but for proportions according to its spatial similitude. This decision faces one trade off. On the one hand, it is harder to choose the best model, on the other hand, it tries to improve chosen model reliability.

\item[Integration] With the integration of spatial bootstrap and Bayesian update, we presented one way of choosing which model is the best, amongst different conceptual models, according to its uncertainty. The best conceptual model is not necessarily the one which reproduces the same structure as the reality, it also depends on the discrepancy of the initial hypothesis in relation to the reality. Such hypothesis is then corrected by the Bayesian update, given the different scenarios considered through spatial bootstrap.
\end{description}
During this work, several algorithms were developed to accomplish every steps described in the presented methodology, using a highly efficient scientific programming language (Fortran), except for the direct sequential simulation, where we used an implementation for the parallelized version \citep{Nunes2010}.

As a side result, we created a tool which allows one to apply the developed methodology in an easy and integrated manner. Thereby, a computational application is now available, which will help on the research and development of this methodology by any student or professional.

\subsection{Recommendations for future work}

In the course of this work, some alternative or additional steps to the developed methodology were emerging, which have all the legitimacy to be developed and tested, however there was a time limitation. Here are some suggestions for a future work.
\begin{description}
\item[Application to real case studies] It would be interesting to apply this methodology to a real case and compare with other methods. Its adaptation and application to other fields, within the Earth sciences, e.g., other natural resources or environmental impact studies, or even other areas of knowledge, would certainly be an interesting challenge.

\item[Drilling strategy] The tested location of wells was, in some way, solely random. Wells which are located in high pay zones should be tested, thus assessing uncertainty when available data is biased.

\item[Scenarios simulations] We just simulated one geological scenario for each pair $(R_k, O_k)$, within the same drilling strategy. Although this number can be enough for what was intended, running several simulations would result in one more measure of uncertainty.

\item[Integrate other data] Seismic information could be incorporated in the spatial bootstrap procedure (e.g., through cosimulation), contributing to the bias reduction related to wells location as well the uncertainty associated to this process \citep{Journel2004}.

\item[Model selector function] The referred approach to select the best model (Section \ref{ss: sensibilidade}) needs a longer time of research and development, and it could prove to be an important tool in reservoir stochastic modeling.
\end{description}
\setlength{\bibsep}{0.5pt}
\bibliographystyle{plain}

\end{multicols}
\end{document}